\documentclass[preprint,12pt,showpacs,aps,amsmath,amssymb, 
nofootinbib,superscriptaddress,hyperref,showkeys]{revtex4} 

\usepackage[utf8]{inputenc}

\usepackage{epsfig} 
\usepackage{wasysym} 
\usepackage{mathrsfs} 
\usepackage{graphicx} 
\usepackage{amsfonts} 
\usepackage{amsbsy} 
\usepackage{amscd} 
\usepackage{pstricks} 
\usepackage{multirow}
\usepackage{slashed} 
\usepackage{tikz}
\usetikzlibrary{decorations.pathmorphing,decorations.markings}
\newcommand{\beq}{\begin{equation}}
\newcommand{\eeq}{\end{equation}}
\newcommand{\bea}{\begin{eqnarray}}
\newcommand{\eea}{\end{eqnarray}}
\newcommand{\beas}{\begin{eqnarray*}}
\newcommand{\eeas}{\end{eqnarray*}}
\newcommand{\bi}{\begin{itemize}}
\newcommand{\ei}{\end{itemize}}

\def\etal{{\it et al.~}}
\def\tev{\,{\ifmmode\mathrm {TeV}\else TeV\fi}}
\def\gev{\,{\ifmmode\mathrm {GeV}\else GeV\fi}}
\def\to{\rightarrow}

\newcommand{\pk}[1]{{\color{red} #1}}

\allowdisplaybreaks

\begin{document}

\author{Selvaganapathy J}
\email{p2012015@goa.bits-pilani.ac.in}
\affiliation{Department of Physics, Birla Institute of Technology and Science-Pilani,
K K Birla Goa campus, Goa-403726, India}

\author{Prasanta Kumar Das}
\email{pdas@goa.bits-pilani.ac.in}
\affiliation{Department of Physics, Birla Institute of Technology and Science-Pilani,
K K Birla Goa campus, Goa-403726, India}

\author{Partha Konar}
\email{konar@prl.res.in}
\affiliation{Physical Research Laboratory, Ahmedabad-380009, India}

\title{Drell-Yan as an avenue to test noncommutative Standard Model at the Large Hadron Collider}

\preprint{\today}

\begin{abstract}
We study the Drell-Yan process at the Large Hadron Collider in presence of the noncommutative extension of standard model. Using the  
Seiberg-Witten map, we calculate the production cross-section to the first order in the noncommutative parameter
$\Theta_{\mu\nu}$. Although this idea is evolving for long, limited amount of phenomenological analysis was done so far and dominantly 
in the context of linear collider. 
Some outstanding feature from this non-minimal noncommutative 
standard model not only modify the couplings over SM production channel, also allow additional nonstandard vertices which can play a significant role. 
Hence in the Drell-Yan process, as studied in present analysis, 
one also needs to account for gluon fusion process at the tree level. Some of the characteristic signatures such as oscillatory azimuthal 
distributions are outcome of the momentum dependent effective couplings.
We explore the noncommutative scale $\Lambda_{NC} \ge 0.4~\rm{TeV}$ considering different machine energy ranging from $7~\rm{TeV}$ to $13~\rm{TeV}$.
\end{abstract}


\maketitle

\section{Introduction}
\label{sec:intro}

The Large Hadron Collider (LHC) has so far been extremely successful in discovering and constraining the 
properties of the last missing bit of the Standard Model (SM) of particle physics, the Higgs boson~\cite{2012gu,2012gk}. 
Apart from some isolated hints, it is broadly evasive lacking any clinching evidence yet from the physics beyond the Standard Model (BSM), 
exploration of which is one of the primary motive for post-Higgs LHC.  
On the other side, it is widely admitted that the SM can at most a very good description for low energy 
effective theory which, in fact, falls short to explain several outstanding issues both in theoretical 
expectations, as well as experimental observations. 

The idea of field theories on the noncommutative (NC) spacetime is rather primeval, yet fascinating by introducing a fundamental length scale in the model consistent with the symmetry \cite{Snyder}. 
These ideas are further revived after realisation of their possible connection with the quantum gravity, where noncommutativity is perceived as an outcome of certain string theory embedded into a background magnetic field \cite{SW}.
Quantum field theory is described by the fields and the local interaction in a continuous space-time point,  
where the canonical position and momentum variables $x_i,~p_j$ are replaced by the operator 
$\hat{x_i},~\hat{p_j}$ which satisfy the commutation relation 
$$\left[\hat{x_i},~\hat{p_j}\right] =i \hbar \delta_{ij}.$$ 
Just like the quantisation in phase space, the spacetime coordinate in the noncommutative spacetime gets replaced by an operator $\hat{x}_\mu$ which satisfy the commutation relation 
\bea
\left[\hat{x}_\mu,\hat{x}_\nu \right]= i\Theta_{\mu\nu} = i\frac{c_{\mu\nu}}{\Lambda_{NC}^2}.
\label{XXTheta}
\eea
where $\Theta_{\mu\nu}$ is an antisymmetric matrix tensor and of dimension $[M]^{-2}$.
Now, one can take out the dimension full part in terms of mass parameter $\Lambda_{NC}~$ and describe it as the fundamental NC scale at which one expects to see the effect of spacetime noncommutativity. 
$c_{\mu\nu}$ is the anti-symmetric constant c-number matrix which gives a preferred directionality and also a non-vanishing contribution results in deviating from exact Lorentz invariance 
in some high energy scale $\Lambda_{NC}$. 
Theoretically this scale is unknown, but one can try to extract the lower bounds directly from the collider experiments by looking at the characteristic signals this framework can provide. 
LEP studied the prediction of the process $e^+ e^- \to \gamma \gamma$ in the noncommutative QED for several orientation
of the OPAL detector and provided the exclusion limit at around 141 GeV \cite{OPAL}. 
With this very moderate bound, we have scope for significant improvement at the future runs of LHC. Drell–Yan process is arguably best explored process at hadron collider. 
With an extremely clean signals of di-lepton, this process is relied upon for calibrating the parton distribution function in hadrons. Any characteristic deviation can 
easily be a basis for the BSM search. 
In our present work, we study this important process in the context of noncommutative framework. 

There are different approaches to study the effect 
of spacetime noncommutativity in a field theory.  One is the Moyal-Weyl (MW) 
approach. In this approach, one replaces the ordinary product between two functions $\phi(x)$ and 
$\psi(x)$ in terms of $\star$ (Moyal-Weyl) product defined by a formal power series expansion of 
\cite{RS,JS,BV}
\begin{equation}
(f \star g)(x)=exp\left(\frac{1}{2}\Theta_{\mu\nu}\partial_{x^\mu}\partial_{y^\nu}\right)f(x)g(y)|_{y=x}.
\label{StarP}
\end{equation}
Here $f(x)$ and $g(x)$ are ordinary functions on $R^n$ and the expansion in the star product can be 
seen intuitively as an expansion of the product in its non-commutativity. 
To describe some of the collider searches of spacetime noncommutativity available in the literature,  Hewett \etal \cite{HPR, Hewett} have studied the processes 
$e^+ e^- \to e^+ e^-$ (Bhabha) and $e^- e^- \to e^- e^-$(Moller) and subsequent studies  \cite{Mathews,Rizzo} were done in the context of 
$e \gamma \to e \gamma$ (Compton) and $e^+ e^- \to \gamma \gamma$ (pair annihilation), $\gamma \gamma \to e^+ e^-$ and 
$\gamma \gamma \to \gamma \gamma$. For a review on NC phenomenology, see \cite{HKM}. In the context of
LHC, the following investigations are in order. 
Noncommutative contribution of neutral vector boson ($\gamma$, $z$) pair production was studied \cite{AOR} at the
LHC and obtained the bound for the NC scale $\Lambda \ge 1~\rm{TeV}$ under some conservative assumptions. 
Further study on the pair producton of charged gauge bosons ($W^\pm$) at the LHC in the noncommutative extension of the standard model found \cite{Ohl} significant 
deviation of the azimuthal distribution(oscillation) from the SM one (which is a flat distribution) for $\Lambda_{NC} = 700~\rm{GeV}$. More recently, 
t-channel single top quark production is calculated at the LHC and found 
significant deviation in cross section can be expected from the Standard Model for $\Lambda_{NC} \ge 980~\rm{GeV}$ \cite{AEN}. 



Second way of dealing this calculation is the Seiberg-Witten approach in which the spacetime noncommutativity is being treated perturbatively via the Seiberg-Witten (SW) map expansion of the fields in terms of noncommutative parameter $\Theta$ \cite{SW}. Here the gauge parameter $\lambda$ and the gauge field $A^\mu$ is expanded as 
\bea \label{swps}
\lambda_\alpha (x,\Theta) &=& \alpha(x) + \Theta^{\mu\nu} \lambda^{(1)}_{\mu\nu}(x;\alpha) + \Theta^{\mu\nu} \Theta^{\eta\sigma} \lambda^{(2)}_{\mu\nu\eta\sigma}(x;\alpha) + \cdot \cdot \cdot \\
A_\rho (x,\Theta) &=& A_\rho(x) + \Theta^{\mu\nu} A^{(1)}_{\mu\nu\rho}(x) + \Theta^{\mu\nu} \Theta^{\eta\sigma} A^{(2)}_{\mu\nu\eta\sigma\rho}(x) + \cdot \cdot \cdot
\eea
The advantage in the SW aprroach over the Weyl-Moyal approach is that it can be applied to any 
gauge theory and matter can be in an arbitrary representation.  Using this SW map Calmet 
\etal first constructed \cite{CJSWW,CW} the {\it minimal} version of the noncommutative standard
model (mNCSM in brief) 
where they derived the ${\mathcal{O}}(\Theta)$ Feynman rules of the standard model interactions 
and found several new interactions which are not present in the standard model. 
All the above analyses were limited to the leading order in $\Theta$.
Phenomenological analysis was carried out for the process $e^+ e^- \to \gamma,Z \to \mu^+ \mu^-$ at the order of $\Theta^2$ and predicted a reach of around $\Lambda_{NC} = 800~\rm{GeV}$ for the NC scale \cite{AMD,AD}. 
successive study \cite{MSD} also focused on the top quark pair production in the NCSM and predicted a similar reach of the NCSM scale. 

The non-commutative standard model essentially the standard model with the background space-time being non-commutative.
Contrary to most other BSM models where particle content and/or the gauge group is extended,
 there is no new massive degree of freedom is included here, but the 
standard model interactions get modified due to space-time noncommutativity. This gives rise 
the modified standard model interaction vertices extending with additional NC contributions. 
Moreover, it also provides a host of new vertices which are absent  in the SM. 
It is demonstrated \cite{CJSWW,CW,Behr} that the minimal version of NCSM can only be realised in 
some definite choice of representation for the traces in gauge field kinetic term. Freedom of this 
choice leads to a more natural extended version.
Melic \etal \cite{MKTSW,Melic} formulated the non-minimal version of NCSM (nmNCSM in brief), 
where the trilinear neutral gauge Boson couplings arise automatically. 
Note that such anomalous vertices were absent in the minimal version and the interactions in fermion 
sector remain unaffected by different choices of representation in gauge action.
Using this formalism associated Higgs Boson production is recently studied associated with the $Z$ boson, 
taking into account of the earth rotation effect in the nmNCSM and found that 
the azimuthal distribution significant differs from the standard model result if the NC scale 
$\Lambda_{NC} \ge 500~\rm{GeV}$ \cite{SDK}.

Exotic new vertices can serve as the Occam's Razor in differentiating the NCSM from other 
new physics models. In this paper, we discuss this possibility considering one of the 
very simple but reliable signature at the hadron collider which can portray its ability by 
distinguishing the effects of space-time noncommutativity from other new physics scenarios. 
Our analysis is based on the parton level Drell-Yan process in producing the lepton pair at
the large hadron collider 
\bea
 p \, p \rightarrow l^+ \, l^- + X
\eea
where light leptons ($l \equiv e, \mu$) of opposite sign is produced at the final state. 
%
What is significant here, besides the standard quark initiated partonic sub-process for 
di-lepton production, gluon initiated processes can also contribute to this production 
cross-section. In the second process new triple gauge boson vertices $K_{Zgg}$ and 
$K_{\gamma gg}$ contribute which arise naturally as an effective vertex in noncommutativity 
although forbidden in the Standard Model.  A study of these vertices have been 
performed by Behr \etal \cite{Behr}.   Using the experimental LEP upper bound from
$\Gamma^{exp}_{Z \to \gamma \gamma} < 1.3 \times 10^{-4}~{\rm GeV}$ and  
$\Gamma^{exp}_{Z \to g g } < 1.0 \times 10^{-3}~{\rm GeV}$,  a correlated bound on 
these vertices were obtained for  NC scale $\Lambda_{NC} = 1~\rm{TeV}$ 
\footnote{Note that, in principle, $K_{Zgg}$ (and $K_{\gamma gg}$) can be zero, in combination 
of other two couplings $K_{\gamma \gamma \gamma}$ and $K_{Z \gamma \gamma}$. But all three 
cannot be zero \cite{Behr} simultaneously and these other two couplings can be tested at the linear collider with 
high degree of precession. We discuss this allowed range in the next section, and also demonstrate 
our results considering different values within this range.}.

 The organization of the paper is as follows. In Sec.~\ref{sec:drellyan}, we describe the 
modified vertices and the new set of vertices (found to be absent in the SM) which may 
potentially contribute to the Drell-Yan lepton pair production at the hadron collider. We next 
obtain the matrix element square for the partonic sub-process and obtain 
the total cross-section, differential cross-section of Drell-Yan lepton pair production. 
In Sec.~\ref{sec:result}, we demonstrate some of the characteristic distributions, such as, lepton pair invariant mass distribution, total 
cross-section, angular distribution. Arguing in favour of them carrying the hallmark for noncommutative effects we probe the sensitivity of the new vertices to the Drell-Yan lepton pair production. 
Finally, in Sec.~\ref{sec:conclusion} we summarize our result and conclude. 

\section{Drell-Yan production in noncommutative standard model}
\label{sec:drellyan}
\begin{figure}[t]
\centering
\includegraphics[width=5.0in,height=1.2in]{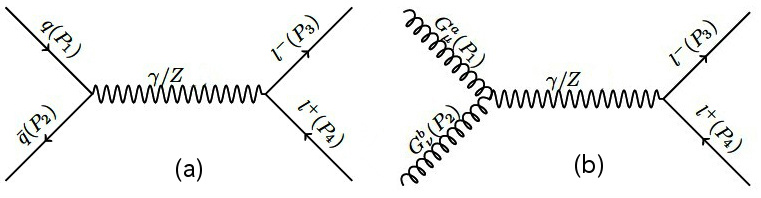} 
  \caption{Representative Feynman diagrams for the partonic subprocess for quark initiated (a) $q \overline{q} \to \gamma, Z \to l^+ l^-$, and gluon initiated (b) $g g \to \gamma, Z \to l^+ l^- $. 
  Both of them contributes in Drell-Yan type lepton pair production at the hadron collider considering
  noncommutative standard model. }
   \label{fig:Figure1}
\end{figure}

\begin{figure}[htb]
\centering
  \includegraphics[width=4in,height=1.2in]{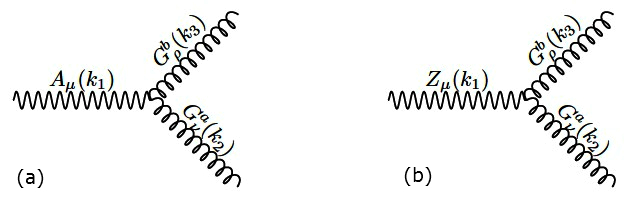}
  \caption{Feynman diagrams for additional vertices in the noncommutative standard model which can contribute in Drell-Yan production process at the LHC.}
   \label{fig:Figure2}
\end{figure}

At the LHC lepton pairs can be produced at the tree-level via the (quark-like) parton initiated 
process (the only partonic level processes possible in the SM at the tree level) 
\begin{eqnarray}
q \overline{q} \to \gamma, Z \to l^+ l^-, 
\end{eqnarray}
together with, in the NCSM additional three boson vertices ensures that lepton pairs can also be 
produced at the tree-level through gluon fusion,
\begin{eqnarray}
g g \to \gamma, Z \to l^+ l^- .
\end{eqnarray}
Representative Feynman diagrams for these partonic subprocess are shown in Fig.~\ref{fig:Figure1}.
For the quark mediated process, the Feynman rules for the vertices 
$f \overline{f} \gamma$ and  $f \overline{f} Z$ (where $f=q,l$) are shown in Appendix \ref{app:feyn}. 
Note that the vertices, besides the SM part, also contain extra $\mathcal{O}(\Lambda_{NC})$ dependent 
term for which, at the limit $\Lambda_{NC} \to \infty$, the original SM vertices get recovered. 
The second gluon mediated partonic-process comprises two new vertices $\gamma g g$ and $Z g g$, 
which are not present in the SM and depicted in Fig.~\ref{fig:Figure2}.  
Corresponding leading order Feynman rules in these figures are given by,
\begin{eqnarray}
 \gamma g g &:&   (-2e) \sin (2\theta_w)  K_{\gamma g g}~  \theta_3^{\mu \nu \rho} (k_1,k_2,k_3)~\delta^{ab} \\  
%
   Z g g  &:& (-2e) \sin (2\theta_w)  K_{Z g g}~  \theta_3^{\mu \nu \rho} (k_1,k_2,k_3)~\delta^{ab}.
\end{eqnarray}
Here, $\theta_W$ is the Weinberg angle and the vertex factors  $K_{\gamma gg}$ and $K_{Zgg}$ are 
given by 
\begin{eqnarray*}
K_{\gamma gg}=\frac{-g_s^2}{2gg'}~(g'^2+g^2)~\zeta_3~, ~~~~ K_{Zgg}=(-\tan \theta_w) ~ K_{\gamma g g},
\end{eqnarray*}
where $g_s,~g,~g'$ are being the $SU(3)_C$, $SU(2)_L$ and $U(1)_Y$ coupling strengths, respectively. 
The tensorial quantity\footnote{We follow the couplings in similar notation as in \cite{Behr} } $\theta_3 \equiv \theta_3^{\mu \nu \rho} (k_1,k_2,k_3)$ and the parameter $\zeta_3$ are defined in Appendix \ref{app:feyn}. 
%
%
 Note that the triple gauge boson vertices $K_{Zgg}$ and 
$K_{\gamma gg}$, absent in the Standard Model (once again, one gets a vanishing $\theta_3$ at the limit $\Lambda_{NC} \to \infty$), arise in this nonminimal version of NCSM. 
A direct test of these vertices have been performed \cite{Behr}  
by studying the SM forbidden decays $Z \to \gamma \gamma$ and $Z \to g g$. 
Analyzing the $3$-dimensional simplex that bounds possible values for the coupling constants $K_{\gamma \gamma \gamma},~K_{Z \gamma \gamma}$ and $K_{Z g g}$ at the 
$M_Z$ scale, allowed region for our necessary couplings $(K_{Zgg}, ~ K_{Z \gamma \gamma})$ are obtained as ranging 
between $(-0.108, -0.340)$ and $(0.217, -0.254)$.

\section{Result and Discussion}
\label{sec:result}


To estimate the noncommutative effects in our parton level calculation, we analytically formulate 
both subprocess initiated either by quark-antiquark pair or by gluon pair at the leading order. 
Using the Feynman rules to $\mathcal{O}{(\Theta)}$ as described above and in Appendix \ref{app:feyn}, the squared 
amplitude (spin-averaged) can be expressed as, 
\begin{eqnarray}
\overline{|M^2_{NCSM}|}_{a b \to l^+ l^-} =  \overline{|\mathcal{M}_{\gamma} + \mathcal{M}_{Z}|^2}   ~~~~ \text{for,} ~~ a,b=q,\bar{q} ~ \text{or} ~ g, g.
\end{eqnarray}
Detailed analytic expression for each nmNCSM amplitude-square is presented in Appendix \ref{app:amp}. 
The NC antisymmetric tensor $\Theta_{\mu\nu}$, analogous to the electro-magnetic field(photon) strength 
tensor, has $6$ independent components: $3$ are of electric type, while $3$ are of magnetic type. We 
have chosen $E_i =  \frac{1}{\sqrt{3}}$ and $B_i = \frac{1}{\sqrt{3}}$ in our analysis. 
(for more, see the Appendix \ref{app:theta}). We have not considered here the effect of earth rotation. 
The impact of earth rotation on $\Theta_{\mu\nu}$ in DY process can be interesting, which is investigated in a future work. \cite{SAPSD}.  
Also note that in the DY lepton distribution, besides the Lorentz invariant momentum dot product ({\sl e.g.}
$p_1.p_2$ etc), the $\Theta$-wieghted  dot product ({\sl e.g.} $p_3 \Theta p_4$ as one follows from Appendix \ref{app:theta}) also appears. 
These terms give rise non-trivial azimuthal distribution in the Drell-Yan lepton pair production as discussed at the end of our results.

%
%
\begin{figure}[t] 
\centering
 \includegraphics[width=3.0in,height=2.0in]{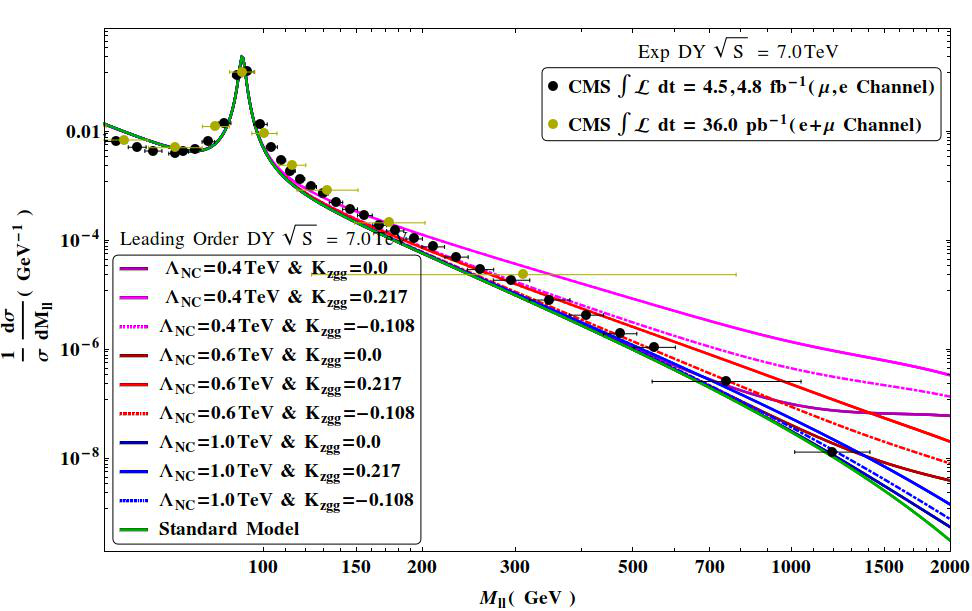}
 \includegraphics[width=3.0in,height=2.0in]{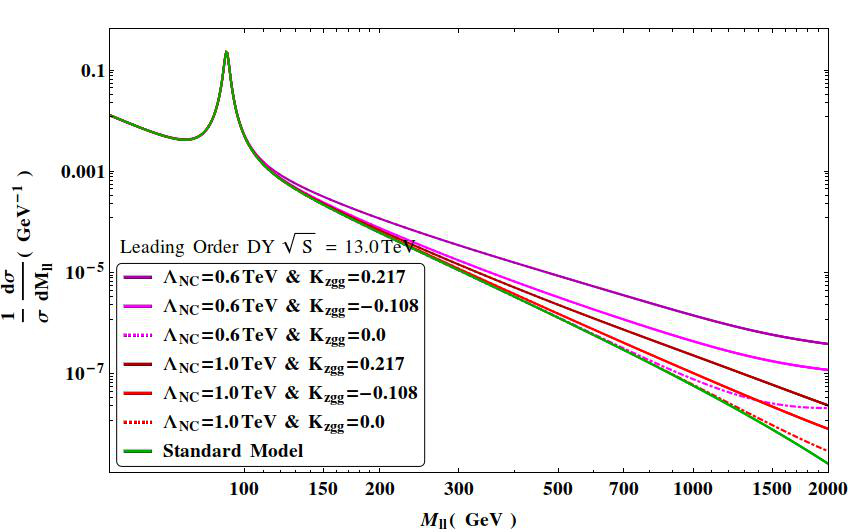}
   \caption{(Color online).  Normalized invariant mass distribution $\frac{1}{\sigma} \frac{d\sigma}{dM_{ll}}~(\rm{GeV^{-1}})$ 
   as a function of the invariant mass $M_{ll}$ (GeV) is shown corresponding to the machine energy 
   (left plot)$\sqrt{s} = 7~\rm{TeV}$  and (right plot) $13~\rm{TeV}$ respectively. Continuous curves 
   of different colours in both plots are shown  for the choice of $\Lambda$ and $K_{Zgg}$ and they 
   converge to the lowermost SM curve in the limit both of these parameters go to zero. In 7 TeV plot, 
   experimental bin-wise data is also shown with central values and error bars. }
   \label{fig:Figure3}
\end{figure} 

We estimate the parton level total cross section and differential distributions for the LHC operated at the energy $\sqrt{S}$,
\begin{eqnarray}
 d\sigma_{pp\to l^+ l^-} =\sum_{ab} \int dx_{1} \int dx_{2} ~ f_{a}(x_{1}, \mu_f^2) f_{b}(x_{2}, \mu_f^2) ~ d\hat{\sigma}_{a b \to l^+ l^-} (x_1 x_2 S).
 \end{eqnarray}
We employ CTEQ6L1 parton distribution function (PDF) throughout the analysis, setting the factorization scale $\mu_f$ at the dilepton invariant mass $M_{ll}$.
%
After formulating the setup, we are now in a position to describe the numerical results for the 
Drell-Yan lepton pair production in the presence of spacetime noncommutativity. In Fig.~\ref{fig:Figure3} we have shown the
normalized  di-lepton invariant mass distribution $\frac{1}{\sigma} \frac{d\sigma}{dM_{ll}}~(\rm{GeV^{-1}})$ against
the invariant mass $M_{ll}$ (GeV) corresponding to the LHC machine energy  $\sqrt{s}$ at  (left plot) $7~\rm{TeV}$ and (right plot) $13~\rm{TeV}$. 
The peak at $M_{ll} = 91.18~\rm{GeV}$ corresponds to the $Z$ boson resonance production.  Different  continuous curves in both plots 
correspond to the theoretical (SM and nmNCSM) predictions. Note that the additional positive contributions in nmNCSM curves 
are realised from two sources; first being the $\Theta$ dependent NC parts supplemented with the SM vertex, and the second being 
complete new tree level process that enhances significantly. 
In the 7 TeV (left) plot the dotted curves correspond to the experimental binwise data provided by CMS collaboration \cite{CMS7}
for the integrated luminosity $4.5~\rm{fb^{-1}}$ and $35.9~\rm{pb^{-1}}$ which is presented along with the error bar.

The lowermost curve in each plot, Fig.~\ref{fig:Figure3} is the SM contribution estimated at the leading order.  In this figure we present different NC contribution based on the 
two relevant parameters $\Lambda_{NC}$ and $K_{Zgg}$ varying between $(0.4~\rm{TeV} - 1~\rm{TeV})$ and (-0.108, +0.217) respectively. Justification for 
these choices are already discussed before. Since the parameter $K_{Zgg}$  contributes in square from gluon initiated diagram, sign of this parameter 
is irrelevant. So, both sign contribute positively and the magnitude depending upon the absolute values. 
%
Note that $K_{Zgg} = 0$ corresponds to the vanishing coupling of gluon with $Z$ and $\gamma$ bosons and 
and the Drell-Yan process in the NCSM arises only from the quark mediated partonic subprocess $q \overline{q} \rightarrow \gamma,Z \rightarrow l^+ l^-$ as in  Figure 1(a). 
The NC scale $\Lambda_{NC}$ determines the energy when this BSM effects can be perceived and this phenomena is evident following different scales in the figure.
At around few hundred of dilepton invariant mass Fig.~\ref{fig:Figure3} exibits, especially at the low $\Lambda_{NC}$ and larger absolute value of $K_{Zgg}$,
 the NCSM effect in this distribution deviating from the SM distribution and it increases with $M_{ll}$.

\begin{figure}[t]
\centering
 \includegraphics[width=3.5in,height=2.4in]{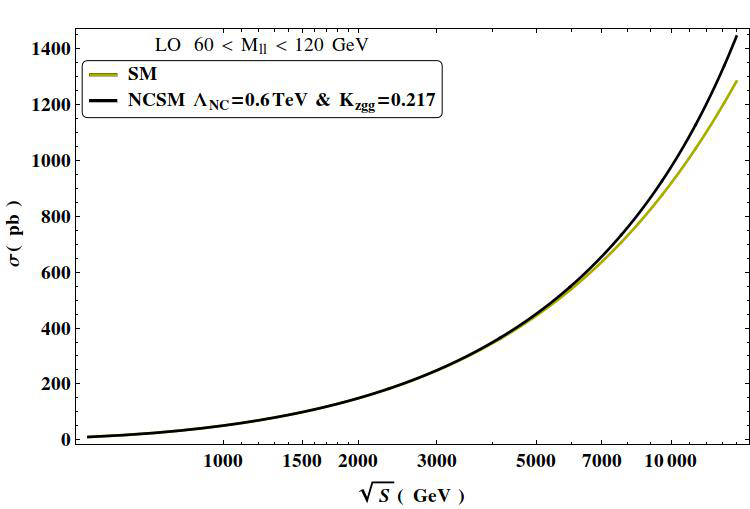}
  \caption{(Color online). The Drell-Yan cross section is shown as a function of the LHC machine 
  energy. In the NCSM, we demonstrate with one of the very optimistic choice like, $\Lambda_{NC}=0.6~\rm{TeV}$ and $K_{Zgg} = 0.217$.}
  \label{fig:Figure4}
\end{figure} 

In Fig.~\ref{fig:Figure4}, we have plotted total leading order Drell-Yan cross section $\sigma$ (in pb) against the LHC collision energy $\sqrt{s}$.
The lower curve corresponds to the SM cross section. We find  $\sigma = 635(1283)~\rm{pb}$ at $\sqrt{s} = 7(14)~\rm{TeV}$, respectively.
To estimate the total cross section, we have considered the di-lepton invariant mass interval 
$60~\rm{GeV} < M_{ll} < 120~\rm{GeV}$. To visualise the effect we once again consider a very optimistic values of  
$\Lambda_{NC} = 0.6~\rm{TeV}$ and $K_{Zgg}=0.217$ for the upper curve corresponds to the NCSM cross section.
For reference we present the corresponding Drell-Yan cross sections for different machine energy in Table \ref{tab:Table1}.
For different machine energy between $7~\rm{TeV}$ and $14~\rm{TeV}$, the leading order SM and the 
nmNCSM (for the same reference parameters) cross sections increases from $636~(656)\rm{pb}$ to $1283~(1438)\rm{pb}$.  
\begin{table}[t] 
 \begin{tabular}{|c|c|c|c|} 
 \hline 
   $ \sqrt{S}$ TeV & $ \sigma_{SM}(pb) $  &  $\sigma_{nmNCSM} (pb)$ & $  \sigma_{EXP} (pb) $ \\
                           &   LO, $\mu_f = M_{z}$     &     LO, $\mu_f = M_{z}$         &             \\  \hline 
   7.0    & 636     &  656             & 974 $ \pm 0.7 $ (Stat) $ \pm 0.7 $ (Syst) \\
   8.0    & 731     &  760             & 1138 $ \pm 8 $ (Stat)                  \\
   13.0  & 1193   &  1325           &                                                                  \\
   14.0  & 1283   &  1438           &                                                     \\
    \hline 
 \end{tabular}
 \caption{ Drell-Yan cross section in the SM, nmNCSM are shown for $ 60~\rm{GeV} < M_{ll} < 120~\rm{GeV} $. The 
experimental data for the same di-lepton invariant mass interval is shown. Here we have set the parameters 
$\Lambda_{NC} = 0.6~\rm{TeV}$ and $K_{Zgg}=0.217$ which is optimistic.}
\label{tab:Table1}
 \end{table} 

\begin{figure}[b]
\centering
 \includegraphics[width=3.5in,height=2.5in]{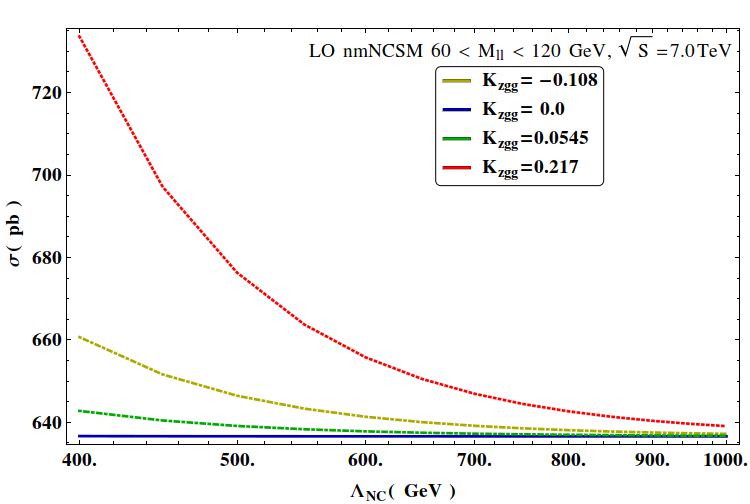}
   \caption{The total cross section for $pp \to (\gamma, Z) \to l^+ l^-$ $ \sigma$ is plotted as a function of the 
   NC scale $\Lambda_{NC}$ (GeV) corresponding to $Z=-0.108,~0.054$ and $0.217$ and fixed machine energy 
   $\sqrt{s} = 7.0~\rm{TeV}$.}
   \label{fig:Figure5}
\end{figure}

In Fig.~\ref{fig:Figure5}, the nmNCSM Drell-Yan cross section is shown as a function of the NC scale $\Lambda_{NC}$ at a fixed machine energy 
$\sqrt{s} = 7.0~\rm{TeV}$. Once again dominant production cross section is estimated for the range of invariant mass $ 60~\rm{GeV} < M_{ll} < 120~\rm{GeV} $ 
corresponding to different values of the parameter  $K_{Zgg}=-0.108,~0.0,~0.0545$ and $0.217$. 
As expected, for a fixed $K_{Zgg}$ coupling  the cross-section $\sigma$ decreases as the NC scale  $\Lambda_{NC}$ increases
and finally merges to the SM value at the very high value of $\Lambda_{NC}$. Note that the NCSM contribution to DY 
process for $K_{Zgg} =0$  almost equal to the SM value as it receives very small
contribution from the quark mediated partonic process and the dominant gluon mediated subprocess is absent 
due to $K_{Zgg} =0$ (and $K_{\gamma gg} =0$). That causes this curve as the lowest (almost) horizontal curve, hence independent of  $\Lambda_{NC}$ scale.
In Table \ref{tab:Table2}  we present the leading order cross sections estimated for different  $\Lambda_{NC} = 0.4,~0.6$ and $1.0~\rm{TeV}$
corresponding to $K_{Zgg} = 0,~-0.108$ and $0.217$. 
\begin{table*}[t]
%
 \begin{tabular}{|c|c|c|c|c|c|c|c|c|} 
 \hline 
    $ \Lambda_{NC} $ (TeV) &  $ K_{Zgg} $ & $ \sigma_{NCSM} $ (pb)  & $ \Lambda_{NC} $ (TeV) &  $ K_{Zgg} $ & $ \sigma_{NCSM} $ (pb)  & $ \Lambda_{NC} $ (TeV) &  $ K_{Zgg} $ & $ \sigma_{NCSM} $ (pb)  \\  \hline 
   0.4  & 0.0  & 637 &0.6   & 0.0 & 637 & 1.0  & 0.0 & 637  \\
   0.4  & -0.108  & 661 &0.6   & -0.108  & 641 & 1.0  & -0.108 & 637 \\
   0.4  & 0.217  & 734 &0.6   & 0.217 & 656 & 1.0  & 0.217 & 639 \\
   \hline
   \end{tabular}
\caption{ Drell-Yan cross section $\sigma(pp \rightarrow l^{+}l^{-})$ in nmNCSM scenario for  the fixed machine energy 
$ \sqrt{s} = 7.0 $ TeV. For $K_{Zgg}=0$, the partonic subprocess $gg \to \gamma, Z \to l^+ l^-$ is absent.}
\label{tab:Table2}
 \end{table*} 



\begin{figure}[t]
\centering
 \includegraphics[width=3.0in,height=2.0in]{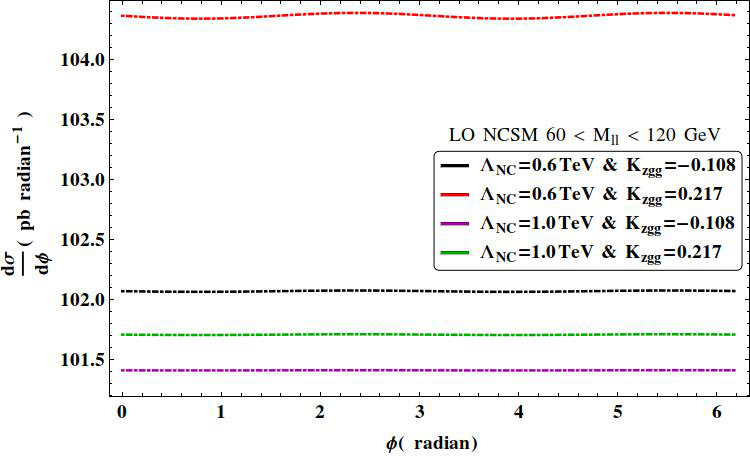} 
 \includegraphics[width=3.0in,height=2.0in]{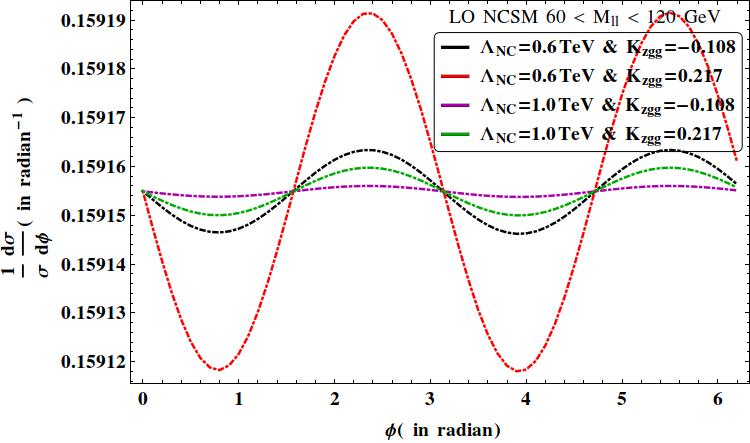}
  \caption{(Color online). $\frac{d\sigma}{d\phi}$ as a function of 
  $\phi$ for $pp \to (\gamma, Z) \to l^+ l^-$ ($l=e,\mu$) for $\Lambda_{NC} = 0.6~\rm{TeV}$, $1.0~\rm{TeV}$ and 
  $K_{Zgg} = -0.108$ and $0.217$, respectively.}
  \label{fig:Figure6}
\end{figure}

After exploring the additional NC contributions coming towards the Drell-Yan production and how different NC parameters can affect 
such processes, now we would like to point out some of the very characteristic distributions attributed to noncommutativity. 
Since space-time noncommutativity essentially breaks the Lorentz invariance which includes the rotational invariance around beam direction,
that can contribute to an anisotropic azimuthal distribution. Angular distributions of the final lepton can thus carry this signature on noncommutativity.
Similar feature is noted in many different process earlier related with the NC phenomenology, nevertheless we would like to present this distribution in our context.
We show the azimuthal angular distribution for the final lepton in Fig.~\ref{fig:Figure6}. 
On the left plot this distribution of the azimuthal angle is shown for  Drell-Yan events if noncommutative effect is there. While the anisotropic effect is not much 
visible here under the considerably large cross-section, it would be evident in the right plot where normalized distribution is demonstrated for that same azimuthal angle.
This figure is generated corresponding to different scales $\Lambda_{NC} ={0.6}~\rm{TeV}$ and $1~\rm{TeV}$. Also, for each $\Lambda_{NC}$, 
we have selected $K_{Zgg} = -0.108$ and $0.217$, respectively.

From the Fig.~\ref{fig:Figure6} right plot, we see that the azimuthal distribution of leptons oscillates over $\phi$, reaching at their maxima at $\phi = 2.342~\rm{rad}$ and 
$5.489~\rm{rad}$. The two intermediate minimas are located at $\phi = 0.783~\rm{rad}$ and $3.931~\rm{rad}$. 
Also note that the azimuthal distribution $\frac{d\sigma}{d\phi}$ is completely flat in the SM. 
A departure from the flat behaviour in the NCSM is due to the term 
$p_4\Theta p_3 (\sim cos\theta + sin\theta(cos\phi + sin\phi))$ term in the azimuthal distribution which brings $\phi$ 
dependence. There still be this feature of azimuthal distribution, even if one deviate from taking the simple form of $\Theta_{\mu \nu}$, however the 
location of peak positions shift.
Such an azimuthal distribution irrespective of peak positions clearly reflects the exclusive nature of spacetime noncommutativity which is 
rarely to be found in other classes of new physics models and can be tested at LHC.   

\section{Summary and Conclusion}
\label{sec:conclusion}
 The idea that spacetime can become noncommutative at high energy have drawn attention after the recent 
 advance in string thery. In this paper we have explored the NC effect in the Drell-Yan lepton pair production
$pp \to (\gamma,Z) \to l^+ l^-$ at the Large Hadron Collider. Two new vertices $Zgg$ and $\gamma gg$ 
(absent in the SM) are being found to play crucial role giving rise new partonic sub-process 
$g g \stackrel{\gamma, Z} {\longrightarrow} l^+ l^-$(absent in the SM).  
For $\sqrt{s} = 7~\rm{TeV}$, as the coupling parameter $K_{Zgg}$ (corresponding to 
the new verices $Zgg$) changes from $-0.108$ to $0.217$, the cross section $\sigma$ increases 
from $637(660)~\rm{pb}$ to  $639(734)~\rm{pb}$ corresponding to $\Lambda_{NC} = 1(0.4)~\rm{TeV}$. 
The azimuthal distribution  $\frac{d\sigma}{d\phi}$, completely $\phi$ independent in the SM, deviates 
substantially in the NCSM. Thus the noncommutative geometry is quite rich in terms of its 
phenomenological implications, which are worthwhile to explore in the TeV scale Large Hadron Collider.


\acknowledgments
The work of P.K.Das is supported in parts by the CSIR Project (Ref. No. 03(1244)12/EMR-II) and the BRNS Project (Ref. No.2011/37P/08/BRNS). 
The authors would like to thank Mr. Atanu Guha for useful discussions. Selvaganapathy J would also like to 
acknowledge his colleague, Mr. Aneesh K.P (recently deceased) for useful discussions.
 


\appendix
\section{Feynman rules}
\label{app:feyn}
\noindent The fermion $f$ (quark $q$ and lepton $l$) coupling to photon and $Z$ bosons, 
to order $\mathcal{O}(\Theta)$ is given by 

\begin{eqnarray}
\gamma f \bar{f} &:& ieQ_{f}\left\{ \gamma_{\mu} + \left[ \frac{i}{2}+\left(\frac{P_{o} \Theta P_{i}}{8} \right)\right] 
 \left[ (P_{o}\Theta)_{\mu} (\Delta \slashed P_{i})+(\Theta P_{i})_{\mu}(\Delta \slashed P_{o}) -(P_{o} \Theta P_{i})\gamma_{\mu}\right] \right\} \\
%
Z f \bar{f} &:& \frac{ieQ_{f}}{sin(2\theta_{W})}\left\{ \gamma_{\mu}\Gamma_{A}^{-}(f) + 
  \left[ \frac{i}{2}+\left(\frac{P_{o} \Theta P_{i}}{8} \right)\right]   \right.  \nonumber \\
   && \;\; \left. \left[ (P_{o}\Theta)_{\mu}(\Delta \slashed P_{i}) \Gamma_{A}^{-}(f) 
  +(\Theta P_{i})_{\mu}(\Delta \slashed P_{o})\Gamma_{A}^{-}(f) 
  -(P_{o} \Theta P_{i})\gamma_{\mu} \Gamma_{A}^{-}(f) \right] \right\}  
\end{eqnarray}
Here we follow the following notation: $i: in,~o:out$ and $\Delta \slashed P_{in,out} = \slashed P_{in,out} - m$. 
Also $ \Gamma_{A}^{-}(f) = c_{V}^f - c_{A}^f\gamma^{5} 
;\hspace{.2in} c_{V}^f = I_{3}^f - 2Q_{f} sin^{2}(\theta_{W}) ; \hspace{.1in} c_{A}^f = I_{3}^f $. 
$Q_f$ is the e.m. charge, and $I_3^f$ is the third component of the week isospin of the fermion $f$ 
(quark($q$) or lepton($l$)).

\noindent The factors $\theta_3$ and $\zeta_3$ arises in $\gamma-g-g$ and $Z-g-g$ are given by 
\cite{Melic},  
\begin{eqnarray}
\theta^{\mu \nu \rho}_{3}(k_1, k_2, k_3) &=& \theta_3[(\mu ,k_1),(\nu ,k_2),(\rho , k_3)]   \nonumber \\ 
& = & -(k_1 \Theta k_2)[(k_1-k_2)^\rho \eta^{\mu \nu}+(k_2-k_3)^\mu \eta^{\nu \rho}+(k_3-k_1)^\nu \eta^{\rho \mu}]   \nonumber \\
& - & \Theta^{\mu \nu} [k_1^\rho(k_2 \cdot k_3)-k_2^\rho (k_1 \cdot k_3)]-\Theta^{\nu \rho} [k_2^\mu(k_3 \cdot k_1)-k_3^\mu(k_2 \cdot k_1)]  \nonumber \\
& -  & \Theta^{\rho \mu} [k_3^\nu(k_1 \cdot k_2)-k_1^\nu(k_3 \cdot k_2)]  +  (\Theta k_2)^\mu [\eta^{\nu \rho}k_3^2-k_3^\nu k_3^\rho] \nonumber \\
& + & (\Theta k_3)^\mu [\eta^{\nu \rho}k_2^2-k_2^\nu k_2^\rho]+(\Theta k_3)^\nu [\eta^{\mu \rho}k_1^2-k_1^\mu k_1^\rho] 
+ (\Theta k_1)^\nu [\eta^{\mu \rho}k_3^2-k_3^\mu k_3^\rho] \nonumber \\
& + & (\Theta k_1)^\rho [\eta^{\mu \nu}k_2^2-k_2^\mu k_2^\nu]+(\Theta k_2)^\rho [\eta^{\mu \nu}k_1^2-k_1^\mu k_1^\nu],  
\end{eqnarray}
and  
\begin{eqnarray}
\zeta_3=\frac{1}{3g_3^2}-\frac{1}{6g_4^2}+\frac{1}{6g_5^2} 
\end{eqnarray}
where $g_3, g_4$ and $g_5$ are the moduli parameters defined in \cite{Behr}.
\section{Squared-amplitude terms}
\label{app:amp}

The amplitude-squared terms for the quark-(anti)-quark initiated partonic sub-process $q \overline{q} \to (\gamma, Z) \to l^+ l^-$ from the Feynman diagram \ref{fig:Figure1}(a),
\begin{eqnarray} 
\overline{|\mathcal{M}_{q,\gamma}|^{2}} = \left( \frac{A F_1}{3} \right) \left[(p_{1}.p_{3})(p_{2}.p_{4})+(p_{1}.p_{4})(p_{2}.p_{3}) \right]
\end{eqnarray}
 \begin{eqnarray} 
 \overline{|\mathcal{M}_{q,Z}|^{2}} & = & \left( \frac{B F_1}{3} \right) (c_{A}^{l^{2}}+c_{V}^{l^{2}})(c_{A}^{q^{2}}+c_{V}^{q^{2}})\left[(p_{1}.p_{3})(p_{2}.p_{4})+(p_{1}.p_{4})(p_{2}.p_{3}) \right] \nonumber \\ 
 & + & \left( \frac{B F_1}{ 3} \right) c_{A}^{l}c_{V}^{l}c_{A}^{q}c_{V}^{q}\left[(p_{1}.p_{4})(p_{2}.p_{3})-(p_{1}.p_{3})(p_{2}.p_{4}) \right]
\end{eqnarray}
and 
 \begin{eqnarray}
  2 Re\overline{|\mathcal{M}_{q,\gamma}|^{\dagger}|\mathcal{M}_{q,Z}|} & = & \left( \frac{ 2C F_1}{3}  \right) c_{A}^{l}c_{A}^{q}\left[(p_{1}.p_{3})(p_{2}.p_{4})+(p_{1}.p_{4})(p_{2}.p_{3}) \right] \nonumber \\ 
 &-& \left(\frac{ 2C F_1 }{3}  \right) c_{V}^{l}c_{V}^{q}\left[(p_{1}.p_{4})(p_{2}.p_{3})-(p_{1}.p_{3})(p_{2}.p_{4}) \right] 
\end{eqnarray}        
 Here $ A = \frac{ 128 \pi^{2} \alpha^{2} Q^{2}_{l} Q^{2}_{q}}{\hat{s}^{2}} $, $ B=  \frac{128 \pi^{2} \alpha^{2} Q^{2}_{l} Q^{2}_{q}}{Sin^{4}2\theta_{W}\left[(\hat{s}-M_{z}^{2})^{2}+(M_{Z}\Gamma_{Z})^{2}\right]} $, $ C=\frac{ 128 \pi^{2} \alpha^{2} Q^{2}_{l} Q^{2}_{q}(\hat{s}-M_z^2)}{\hat{s}[(\hat{s}-M_z^2)^2 + M^2_z \Gamma^2_z]}$
  , $ F_1 =  \left[ 1+ \frac{\left(P_{2}\Theta P_{1}\right)^{2}}{4}\right] \left[ 1+ \frac{\left(P_{4}\Theta P_{3}\right)^{2}}{4}\right]$ and 
  $\hat{s}=s x_{1}x_{2}$. Note that ampltude-square terms goes as 
  $\mathcal{O}\left( 1, \frac{1}{\Lambda^2_{NC}},  \frac{1}{\Lambda^4_{NC}}\right)$, respectively. \\

\noindent For the gluon initiated partonic sub-process $g  g \to (\gamma, Z) \to l^+ l^-$, the squared-amplitude terms from the Feynman diagram \ref{fig:Figure1}(b),
are given by 
\begin{eqnarray}
 \overline{|\mathcal{M}_{g,\gamma}|^{2}}= 4 D F_2 \left(p_{3 \rho} p_{4 \sigma}+p_{3 \sigma} p_{4 \rho}-\eta_{\rho \sigma} p_{3} \cdot p_{4} \right)
 \cdot \left( \eta_{\nu \beta} \bar{\theta_3}^{\alpha \beta \sigma} \eta_{\alpha \mu} \bar{\theta_3}^{\mu \nu \rho } \right)
\end{eqnarray}
Similarly,
\begin{eqnarray}
 \overline{|\mathcal{M}_{g,Z}|^{2}} & = &  4 G F_2 (c_{A}^{l^{2}}+c_{V}^{l^{2}}) \left(p_{3 \rho} p_{4 \sigma}+p_{3 \sigma} p_{4 \rho}-\eta_{\rho \sigma} p_{3} \cdot p_{4} \right) \left( \eta_{\nu \beta} \bar{\theta_3}^{\alpha \beta \sigma} \eta_{\alpha \mu} \bar{\theta_3}^{\mu \nu \rho } \right)
\end{eqnarray}

\begin{eqnarray}
 2 Re \overline{|\mathcal{M}_{g,\gamma} \mathcal{M}^\dagger_z|} & = & 4 H F_2 
  c_{V}^{l}\left(p_{3 \rho} p_{4 \sigma}+p_{3 \sigma} p_{4 \rho}-\eta_{\rho \sigma} p_{3} \cdot p_{4} \right)  \cdot  \left( \eta_{\nu \beta} \bar{\theta_3}^{\alpha \beta \sigma} \eta_{\alpha \mu} \bar{\theta_3}^{\mu \nu \rho } \right) 
\end{eqnarray}

\noindent Here $ F_2=  \left[ 1+ \frac{\left(P_{4}\Theta P_{3}\right)^{2}}{4}\right]$, $ D = 2 \left( \frac{\pi \alpha \sin (2 \theta_w)K_{\gamma gg}}{\hat{s}} \right)^2$, 
 $G = \left[\frac{2 \pi^2 \alpha^2 k^2_{z gg}}{[(\hat{s}-M_z^2)^2+ M^2_z \Gamma^2_z]}\right]$, \\
 $H= \sin (2 \theta_w) K_{ \gamma gg} K_{zgg} \left[ \frac{4 \pi^2 \alpha^2}{\hat{s}} \left(\frac{\hat{s}-M_z^2}{[(\hat{s}-M_z^2)^2 + M^2_z \Gamma^2_z]} \right) \right]$. \\ \\
The quantity $\bar{\theta}_{3}$ appearing in several squared-amplitude terms, is given by
 \begin{eqnarray*}
   \bar{\theta}_{3}^{\mu \nu \rho}  &=&  -(p_1 \Theta p_2)[(p_1-p_2)^\rho \eta^{\mu \nu}+2(p_2^\mu \eta^{\nu \rho}-p_1^\nu \eta^{\rho \mu})] \\
                    & + & (p_1 \cdot p_2)[\Theta^{\mu \nu}(p_1-p_2)^\rho-2((p_2\Theta)^{\mu}\eta^{\nu \rho}+(p_1\Theta)^{\nu}\eta^{\mu \rho})]  + [(p_2 \Theta)^\mu p_1^\nu+(p_1 \Theta)^\nu p_2^\mu](p_1+p_2)^\rho 
 \end{eqnarray*}
 Note that ampltude-square terms goes as 
  $\mathcal{O}\left( \frac{1}{\Lambda^4_{NC}},  \frac{1}{\Lambda^8_{NC}}\right)$, respectively. \\
 \noindent In evaluating the matrix element square, we have used the following orthonormality condition 
  \begin{eqnarray}
  \sum_{\lambda,\lambda'} \epsilon_{\mu'}^{*a'}(p_1,\lambda_1') \epsilon_\mu^a(p_1,\lambda_1) & = & -\eta_{\mu' \mu} \delta_{a'a} \\
  \sum_{\lambda,\lambda'} \epsilon_\nu^b(p_2,\lambda_2) \epsilon_{\nu'}^{*b'}(p_2,\lambda_2') & = & -\eta_{\nu' \nu} \delta_{b'b} 
  \end{eqnarray}
and the color algebra $ \sum_{aa'bb'}\delta_{bb'} \delta_{aa'} \delta^{ab} \delta_{a'b'}= \sum_{ab} \delta^{ab} \delta_{ab}=\sum^8_{a=1}\delta_{aa}=8 $.
\section{Antisymmetric tensor $\Theta_{\mu\nu}$ and $\Theta$ wieghted dot product}
\label{app:theta}
The antisymmetric tensor $\Theta_{\mu\nu} = \frac{1}{\Lambda_{NC}^2} c_{\mu\nu} $ has $6$ independent 
components corresponding to $c_{\mu\nu} = (c_{oi}, c_{ij})$ with $i,j = 1,2,3$. Assuming them as the 
non-vanishing components, we can write them as follows: 
\begin{eqnarray}
c_{oi} = \xi_i, ~~
c_{ij} = \epsilon_{ijk} \chi^k
 \end{eqnarray}
 
The NC antisymmetric tensor $\Theta_{\mu\nu}$ is analogous to the electro-magnetic(e.m.) field 
strength tensor $F_{\mu\nu}$: $\xi_i$ and $\chi_i$ are like the electric and magnetic field vectors. Setting 
$\xi_i = (\vec{E})_i = \frac{1}{\sqrt{3}}$ and $\chi_i = (\vec{B})_i = \frac{1}{\sqrt{3}}$ 
with $i=1,2,3$ and noting the fact that $\xi_i = - \xi^i,~\chi_i = - \chi^i$, the normalization condition
$\xi_i \xi^j = \frac{1}{3} \delta^j_i$ and  $\chi_i \chi^j = \frac{1}{3} \delta^j_i$, we may write 
$\Theta_{\mu\nu}$ as  
\begin{equation}
 \Theta_{\mu \nu} = \frac{1}{\sqrt{3} \Lambda_{NC}^{2}}\left( \begin{array}{cccc}
0 & 1 & 1 & 1   \\
-1 & 0 & -1 & 1 \\
-1 & 1 & 0 & -1 \\
-1 & -1 & 1 & 0 \\
\end{array} \right)
\end{equation}
Using these we may write the $\Theta$-weighted dot-product as follows:
  \begin{eqnarray}
 p_2 \Theta p_1 &=& \frac{\hat{s}}{2 \sqrt{3} \Lambda_{NC}^2} \\
 p_4 \Theta p_3 &=& \frac{\hat{s}}{2 \sqrt{3} \Lambda_{NC}^2} \left[ cos\theta + sin\theta (cos\phi + sin\phi)\right]
  \end{eqnarray}
  
\noindent Here we have not considered the effect of earth rotation on the anti-symmetric tensor  $\Theta_{\mu\nu}$ in DY process. This
will be reported elsewhere \cite{SAPSD}.




\end{document}